# Human Culture : A History Irrelevant and Predictable Experience

Hao Wang
Haow85@live.com
Ratidar.com
Beijing, China

**Abstract**

Human culture research has witnessed an opportunity of revolution thanks to the big data and social network revolution. Websites such as Douban.com, Goodreads.com, Pandora and IMDB become the new gold mine for cultural researchers. In 2021 and 2022, the author of this paper invented 2 data-free recommender systems for AI cold-start problem. The algorithms can recommend cultural and commercial products to users without reference to users' past preferences. The social implications of the new inventions are human cultural tastes can be predicted very precisely without any information related to human individuals. In this paper, we analyze the AI technologies and its cultural implications together with other AI algorithms. We show that human culture is (mostly) a history irrelevant and predictable experience.

**Introduction**

During the past centuries, one of the difficulties of cultural studies research is the difficulty to collect large data samples for investigation. Without modern technologies , it was impossible to carry through research study on millions of people or billions of people. Thanks to the advent of Web 2.0, websites such as Facebook, Twitter, Douban, Goodreads, IMDB, etc. have collected a huge amount of online data. These datasets contain details of people's online behavior, online remarks, e-commerce purchase logs, cultural ratings of cultural products like movies and music. In addition to the availability of large-scale human data, modern computer technologies such as data mining and artificial intelligence have emerged as major engines for both commercial products and academic research.

One of the data mining fields that has evolved for decades is recommender systems. A recommender system, in plain words, is a data mining / artificial intelligence product that recommend interesting items to users (most commonly) based on their online data. Such online data usually include users' online behavior or demographic / financial information. Examples of recommender systems include Amazon's "Product related to this item", "Recommended based on your shopping trends", and "Customers who viewed items in your browsing history also viewed", etc. are all recommendation functionalities and product lines, and also on Douban.com (a Chinese social network website similar to Goodreads), we have "Guess what you'd like to read", "Guess what you'd like to watch", and "Guesss what you'd like to listen" product features.

The earliest forms of recommender systems technologies are mostly shallow models, which are simple but intepretable. For example, collaborative filtering [1][2] and matrix factorization [3]. Both of these two frameworks have many different variants, but many of them do not take account into the time information related to the items or users that they recommend and the result, if measured in metrics of Mean Absolute Error or Rooted Mean Squared Error, are fairly accurate, and the room for further improvement is really very little.

Since time-free recommender systems are also very successful on cultural websites such as Goodreads, TikTok and Douban, we switch the technical perspective of recommender system to a cultural one. We notice that the implication of the success of time-free recommender systems is that human culture is mostly a collective experiences that are time invariant. Or, in other words, our cultural taste is shaped by our historic preference, but not so much by when we experience the cultural products. For example, if we read *Harry Porter* in 21 and *The Three Body Problem* in 24, it wouldn't be so different from reading *Harry Porter* in 24 and *The Three Body Problem* in 21, when it comes to the shaping of our cultural taste in the future.

In 2021 and 2022, the author of this paper invented 4 data-free recommender system algorithms for the cold-start problem of recommender system. The so-called cold-start problem refers to the phenomenon that when a new user arrives in the website or when a new item is put online, there is no historic information to start the recommendation process. Usually researchers use technologies such as *Transfer Learning [4]* and *Meta Learning [5]* to tackle the problem. However, both paradigms require extra data input sources, which are usually not available in common commercial environments.

The 4 algorithms invented by the author are the first batch of cold-start solvers in history that require no data at all and achieves competitive results with technologies that do use historic data for comparison. The names of the algorithms are ZeroMat [6], DotMat [7], PoissonMat [8] and PowerMat [9]. The first 3 algorithms are context-free and data-free algorithms while PowerMat is context-aware recommendation approach that uses contextual information but nothing else.

Once again, on cultural websites, these algorithms produce strikingly good performance compared with classic recommender system approaches that do use historic data. The implication is that human culture is predictable fairly accurately on individual level without past data. The conclusion is surprising (or horrific) that human culture is history-irrelevant and predictable, and after some time of evolution, evolves into a lock-state [10]. The major reason behind this phenomenon is because of a probability distribution named Pareto Distribution (Power Law Distribution / Zipf Distribution).

In this paper, we will fully analyze the aforementioned recommender system technologies and their cultural implications in detail. We believe our research study marks the beginning of a new field that we'd like to call *Computational Cultural Studies*. We wish our work could point to new and meaningful research directions for scholars and cultural industry workers.

**Related Work**

As a cultural research toolkit, recommender system is comparatively new. However, its application in cultural and social industries has a very long history. Earliest applications of recommendation ideas could be dated back to 1990's or even 1980's. The recommender system has long be viewed as an ideal substitute for search engine and new method for information seeking for the Web 2.0 age.

Recommender systems are widely applied in cultural and social websites. TikTok, Pandora, Douban and Goodreads all embraced recommender system at their very early development age. A survey on recommender technologies, therefore helps us better understand our research topic in this paper since our major focus is on data acquired by such websites, and our major research toolkit is recommender system and its cultural analysis.

Collaborative filtering algorithm and its variants [1][2] are considered as the first batch of inventions of recommender systems. The basic ideas behind the paradigm is to recommend according to references to other similar people's preferences. The algorithm used to be so successful, that even today, it is still widely adopted in the industry as the first matching step to a bigger hybrid model.

Linear models used to be extremely popular in the industry. Internet companies such as Netease [11] and Baidu [12][13] have used linear models, or linear hybrid models to solve dating or question-answer recommendation problems. The major advantages of linear models are its simplicity to implement, extensibility by feature engineering, and efficiency.

Matrix factorization is also a very popular shallow model used in recommender system domain. One of the earliest matrix factorization model used in practice is SVD, which was introduced in the Netflix One Million Dollar Competition. Variants of matrix factorization such as Alternating Least Squares [14], SVD++ [15], SVDFeature [16] were common names in the industry. In recent years, researchers have started to use matrix factorization variants to tackle problems other than technical accuracy. MatRec [17], Zipf Matrix Factorization [18], Focused Learning [19], and KL-Mat [20] were proposed to solve the fairness problem, while MatMat [21], MovieMat [22], and PowerMat [9] were introduced for context aware recommendation.

Learning to rank as a new technique for recommender system problem emerged around 2010. Different from other recommender system, which aim to solve for accuracy problem, the paradigm models the recommendation problem as a ranking optimization problem. Bayesian Personalization Ranking algorithm [23] optimizes for pair-wise ordering, and Collaborative Less is More Filtering [24] is a technique for list-wise optimization. In 2022, a new pairwise learning to rank algorithm named Pareto Pairwise Ranking [25] was introduced for solving fairness problems, following the latest trend of fair learning to rank in both the recommendation and other information retrieval domain [26][27][28].

Deep learning approaches emerged shortly after 2010 as a major technical backbone for most recommender system products. Important deep learning inventions such as Deep Matrix Factorization [29], DLRM [30], DeepFM [31], Wide&Deep [32], AutoInt [33] have attracted attention from many academic researchers and industrial practitioners. The reason for the emergence of deep models is because they can represent a broader range of functions from which the algorithmic designers can choose to fit the data more accurately than a specific function chosen for shallow models.

Recommender systems have not raised attention from the cultural researchers until in 2022, when Wang published a paper titled *Is Human Culture Locked by Evolution?* [10] In this paper, Wang discussed a cold start solution he invented, namely ZeroMat [6], and its cultural implications. ZeroMat was the first Zeroshot recommender systems ever invented in the history, followed by DotMat [7], PoissonMat [8] and PowerMat [9].

**Recommender Systems**

Before we delve into the technical details of recommender systems, we first introduce the datasets that we use for our research study. We use MovieLens Dataset [34] and LDOS-CoMoDa Dataset [35] as our major test benchmarks. MovieLens Dataset is provided by GroupLens. It is an open dataset collections that contains many datasets such as MovieLens Small Dataset, MovieLens 1 Million Dataset, etc. The dataset collection contains users and their ratings of watched movies. LDOS-CoMoDa is another open data set containing user movie watching rating data, but also with the contextual information such as location, mood, etc. during movie watching.

Recommender systems have many different algorithmic paradigms. The major research focus of this paper are the following algorithms: *Collaborative Filtering* , *Matrix Factorization*, *ZeroMat*, *DotMat* and *PoissonMat*. The first 2 algorithmic framworks were early inventions in the history of recommender systems before the year of 2010. Although a bit out-of-dated, collaborative filtering is still widely used as the first stage of many industrial recommender systems. This stage is called matching, followed by a stage named ranking. The stage of matching aims to reduce the set of candidate recommendation set for further processing in the ranking stage.

We take a closer look at item-based collaborative filtering (invented in 2001) . The prediction of user item rating of user u on item i is defined as follows [1] :

$$P_{u,i} = \frac{\sum_{\text{all similar items, N}}(s_{i,N} * R_{u,N})}{\sum_{\text{all similar items, N}}(|s_{i,N}|)}$$

where s denotes the similarity score between two items, and R is the user item rating value. Authors of the original item-based collaborative filtering paper uses Mean Absolute Error (MAE) as the evaluation metric for the algorithm. MAE is defined below :

$$\text{MAE} = \frac{1}{N}\frac{1}{M}\sum_{u=1}^{N}\sum_{i=1}^{M}|P_{u,i} - R_{u,i}|$$

where u iterates through the user set, i the item set. P denotes the predicted user item rating value, and R represents the true values of user item rating in the test data set. The MAE scores of different variants of item-based collaborative filtering on MovieLens Dataset is illustrated in Fig.1 in the original paper of item-based collaborative filtering.

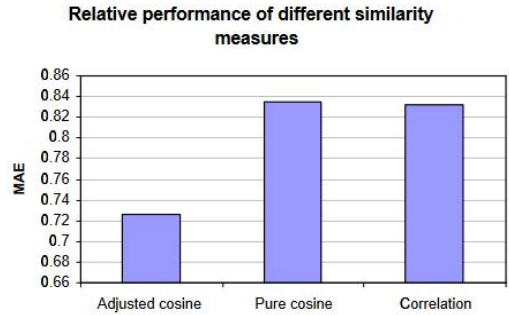

Fig. 1 MAE of Item-based Collaborative Filitering Algorithms ([1])

The maximum user item rating value of MovieLens Dataset is 5. By observation, item-based collaborative filtering variants are *time-order-invariant*, and the prediction is fairly accurate. By *time-order-invariant,* we mean the order of user's past behavior is irrelevant to the algorithmic results. We will discuss the cultural implications of this in the following section.

Matrix Factorization is another popular technique used for recommendation in the past 10 years. The advent of matrix factorization is pretty early, but the first rigorous probabilistic theory supporting the framework was published in 2007. The most classic loss function of matrix factorization is defined as follows :

$$L = \sum_{i=1}^{n} \sum_{j=1}^{m} (R_{i,j} - U_i^T \cdot V_j)^2$$

The basic idea behind matrix factorization is to use the dot products of lower dimensional vectors to fill in blank values of higher dimensional matrices. It solves the accuracy problem of recommender system, while at the same time serving as a dimensionality reduction technique. The algorithm is solved using a standard optimization technique named Stochastic Gradient Descent (SGD). For quite a long time, different variants of the matrix factorization algorithms dominate the landscape of recommendation products.

As noted from the loss functions of the classic matrix factorization and one of its most successful variants Alternating Least Squares. The algorithms are also time-order-invariant. There have been later inventions and improvements on the classic version of the formulation that does take into account of time-order, but similar to collaborative filtering, the space for improvement is slim.

We delve into the details of the 4 cold-start problem solutions now. ZeroMat is a variant of matrix factorization framework, that takes advantage of the observation that user item rating following power law distribution. To put it in plain language, the movie that receives the 5 star rating has the number of ratings proportional to value 5, the movie that receives the 3 star rating has the number of ratings proportional to value 3, etc. This is compliant with the observation that the most popular movies such as Harry Porter are highly praised by a disproportionally more number of people than less popular movies. The basic idea behind ZeroMat is to replace the normal distribution definitions in the classic probabilistic matrix factorization framework by Zipf's Law distribution, as we discussed above.

Formally, we assume that :

$$\frac{R_{i,j}}{R_{max}} \sim \frac{U_i^T \cdot V_j}{max(U_i \cdot V_j)}$$

By this assumption, the loss function of the classic probabilistic matrix factorization becomes :

$$P(U, V \mid R, \sigma_U, \sigma_V) = \prod_{i=1}^{N}\prod_{j=1}^{M}\left(U_i^T \cdot V_j\right) \times \prod_{i=1}^{N} e^{-\frac{U_i^T \cdot U_i}{2\sigma_U^2}} \times \prod_{j=1}^{M} e^{-\frac{V_j^T \cdot V_j}{2\sigma_V^2}}$$

Just like matrix factorization, we apply Stochastic Gradient Descent (SGD) for solving the optimal values of U and V in the log of the MAP, and acquire the following update rules (We also set the standard deviations to be constant value 1) :

$$U_i = U_i + \gamma\left(\frac{V_j}{U_i^T \cdot V_j} - 2 \times U_i\right)$$

$$V_j = V_j + \gamma\left(\frac{U_i}{U_i^T \cdot V_j} - 2 \times V_j\right)$$

Since the SGD formulas do not involve the values of user item rating values R, we are safe to draw the conclusion that the ZeroMat algorithm is a data-free algorithm that is a perfect ideal for the cold-start problem. Actually in recent years, there emerged a field of AI research named *Zeroshot-Learning* that endeavours to solve AI problems without using any data. To the best of our knowledge, ZeroMat is the first Zeroshot recommender system algorithm.

When tested on MovieLens 1 Million Dataset, the author obtained the following results :

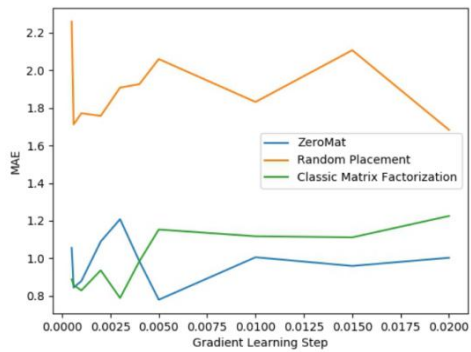

Fig. 2 MAE comparison between ZeroMat and other algorithms ([6])

Notice that when compared with the classic matrix factorization with full historic dataset, the algorithm produces competitive and even (sometimes) superior performance. The cultural implications of this algorithm can never be over-emphasized, as we shall discuss in the following section.

In the following year (2022) of the publication of ZeroMat algorithm, DotMat was proposed by the same author. The algorithm performs even better than ZeroMat. It was built upon the idea of RankMat and Power Law Distribution. RankMat is an algorithm that proposes using Power Law family instead of Exponential Family. The loss function of DotMat is, therefore, after borrowing the idea from RankMat, defined as follows :

$$L = |(U_i^T \cdot V_j)^{U_i^T \cdot v_j} - \frac{R_{i,j}}{R_{max}}|$$

Once again, we apply SGD technique to solve for the optimal values of U and V, and obtain the following formulas :

$$U_i = U_i - \gamma((U_i^T \cdot V_j)^{U_i^T \cdot V_j} \mathrm{sign}((U_i^T \cdot V_j)^{U_i^T \cdot V_j} - \frac{R_{i,j}}{R_{max}})(1 + \log(U_i^T \cdot V_j))V_j)$$

$$V_j = V_j - \gamma((U_i^T \cdot V_j)^{U_i^T \cdot V_j} \mathrm{sign}((U_i^T \cdot V_j)^{U_i^T \cdot V_j} - \frac{R_{i,j}}{R_{max}})(1 + \log(U_i^T \cdot V_j))U_i)$$

A simplified version of the SGD formulas are listed below :

$$U_i = U_i - \gamma((U_i^T \cdot V_j)^{U_i^T \cdot V_j} \mathrm{sign}((U_i^T \cdot V_j)U_i^T \cdot V_j - U_i^T \cdot V_j)(1 + \log(U_i^T \cdot V_j))V_j)$$

$$V_j = V_j - \gamma((U_i^T \cdot V_j)^{U_i^T \cdot V_j} \mathrm{sign}((U_i^T \cdot V_j)U_i^T \cdot V_j - U_i^T \cdot V_j)(1 + \log(U_i^T \cdot V_j))U_i)$$

Careful observations of the SGD formulas lead to the conclusion that DotMat does not rely on user item rating values R, and therefore is a data-free recommendation algorithm. The performance of DotMat is actually superor to ZeroMat, and both of these algorithms could be used as a sparsity mitigation preprocessing step to a hybrid model. The experimental results of DotMat compared with other algorithms are illustrated below :

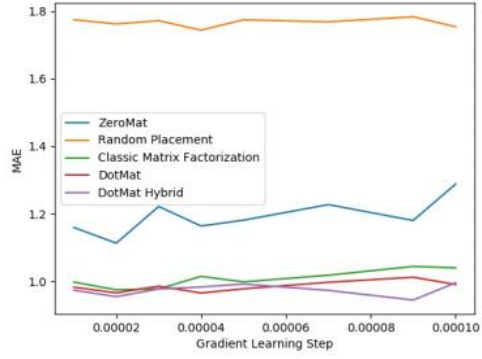

Fig. 3 MAE comparison among DotMat and other algorithms ([7])

Instead of resorting to Zipf distribution, in 2022 a new algorithm named PoissonMat was introduced to replace the Zipf distribution with Poisson distribution. Poisson distribution, as well known in the research circle, represents the probability of the rare event occurrences. The framework behind PoissonMat is also the probabilistic matrix factorization :

$$P(U, V \mid R, \sigma_U, \sigma_V) \sim \frac{\left(\frac{1}{n}\sum_{j=1}^{n} U_i^T \cdot V_j\right)^{U_i^T \cdot V_j} e^{-\frac{1}{n}\sum_{j=1}^{n} U_i^T \cdot V_j}}{\left(\frac{1}{rank}\right)!} \left(U_i^T \cdot V_j\right)$$

To solve for the optimal values of U and V, we take the natural log of the posterior distribution and apply SGD rules :

$$\frac{\partial L}{\partial U_i} = \left(\frac{(U_i^T \cdot V_j + 1)}{U_i^T \cdot V_j} + log(U_i^T \cdot V_j) - 1\right) V_j$$
$$\frac{\partial L}{\partial V_j} = \left(\frac{(U_i^T \cdot V_j + 1)}{U_i^T \cdot V_j} + log(U_i^T \cdot V_j) - 1\right) U_i$$

It is apparent from the update rules that PoissonMat is data-free. Once again, PoissonMat can also serve as a sparsity mitigation preprocessing step to a hybrid model. The experimental result is illustrated below :

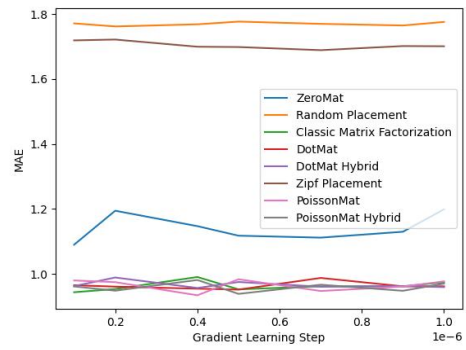

Fig.4 MAE Comparison among PoissonMat and others on MovieLens Dataset ([8])

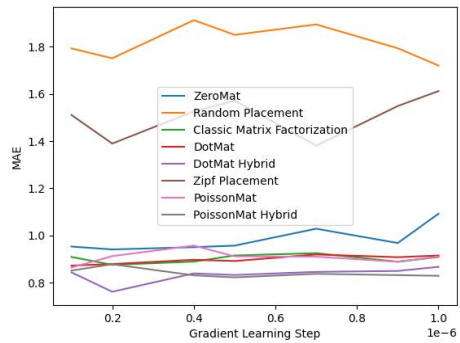

Fig. 5 MAE Comparison among PoissonMat and others on LDOS-CoMoDa Dataset ([8])

PoissonMat is competitve with DotMat while PoissonMat Hybrid is the best performing algorithm in comparison with other approaches.

The last Zeroshot learning algorithm we discuss in this section is PowerMat. PowerMat is a data-free but context-aware algorithm that relies only on contextual information to solve the context-aware recommendation (CARS) problem.

The loss function of PowerMat is defined as follows :

$$P(U, V \mid R, c) = (U \cdot V)^{\alpha \cdot c + \beta U \cdot V}$$

which is a much simplified version of the probabilistic matrix factorization paradigm. The Stochastic Gradient Descent (SGD) rules for this algorithm is defined as follows :

$$U = U - \gamma(\beta U^T \cdot V \cdot V + (\beta U^T \cdot V + \alpha \cdot c)V - \frac{2}{\sigma_U}U)$$

$$V = V - \gamma(\beta V^T \cdot U \cdot U + (\beta V^T \cdot U + c^T \cdot \alpha)U - \frac{2}{\sigma_V}V)$$

$$\alpha = \alpha - \gamma(U^T \cdot V \times c)$$

$$\beta = \beta - \gamma((U^T \cdot V)^2)$$

The update rules utilize only the contextual information vector c, rather than the full historic data matrix R. PowerMat is a data-free but context-aware algorithm that deals whose superiority is demonstrated in the following figure :

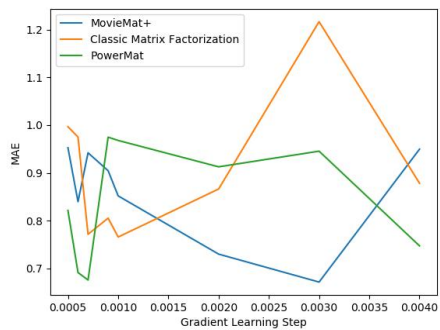

Fig. 6 MAE comparison among PowerMat and other algorithms ([9])

PowerMat is competitive with another movie CARS algorithm - MovieMat+ on MAE scores when compared on the LDOS-CoMoDa Dataset.

In this section, we have overviewed 6 recommender system paradigms / approaches, namely collaborative filtering, matrix factorization, ZeroMat, DotMat, PoissonMat and PowerMat. We analyzed the technical details of the 6 recommender systems and illustrated the superiority of each paradigm / approach in predicting human cultural tastes on individualistic levels. In the next section, we will analyze the cultural implications of these algorithms, and draw several important conclusions.

**Cultural Implications**

Both MovieLens and LDOS-CoMoDa datasets are large-scale cultural datasets that have been extensively researched by data mining professionals, but largely been overlooked by cultural researchers. The technical aspects of recommender systems tested on the two benchmarks have been published in tens of thousands of literature works. We now analyze the cultural implications of these results, exemplified by the 6 paradigms / approaches we discussed in the previous section.

We first discuss the implications of time-order-invariant properties of collaborative filtering and matrix factorization algorithms. The technical accuracy of the two paradigms are illustrated in the previous section. Now consider we are facing the following problem : We meet a new person and we have his historic preference data for movies, how can we guess what he'd like to watch from the movie candidate pool we have by hand ? A simple way is to guess randomly by picking up a random movie. If the maximum rating star of the movies is 5, probably we'll get an MAE of 2.5 if we use this strategy. Actually the error metric is close to 1.8 in our experiments shown in Fig. Xxx to Fig. YYY.

Can we do better than random guess ? Item-based collaborative filtering and matrix factorization, by careful tuning, can reduce the MAE to 0.7 to 0.8. This means that for an item, our guessed user preference would not deviate by more than 1 star from the ground truth. This is a strikingly good result, since most people would not distinguish too much between a 3 star movie and a 4 star movie. Since the MovieLens and LDOS-CoMoDa Datasets take on only integer movie rating values, further improvement on MAE that reduces the values from 0.7 to, let's say, 0.5, would not be so perceptible on the integral scale of the rating values.

Both collaborative filtering and matrix factorization techniques are time-order-invariant. If we take a close look at the technical details of the two algorithmic paradigms, we'll notice that although the algorithms rely on users' past preferences over a set of items. But the time-order of the set of items is irrelevant to the algorithms. ***The cultural implications of this phenomenon is that the time-order of cultural consumption is irrelevant to users' future choice.***

This conclusion might explain why there are global-wise phenomenal cultural products. If time-order does play a vital role in people's choice of cultural products, due to the high diversity of the global population and cultural market, it would be extremely difficult to create a movie that can win the hearts of tens of millions of people. However, such phenomenal cultural products do exist in real world such as Hollywood's blockbusters. This can be explained by our conclusion that time-order of cultural consumption is irrelevant.

Each person's cultural consumption is a time series, given the large population of human beings and the great diversity of cultural market, the diversity of person's cultural preference sequence is extremely high. Let's take the movie market as an example. Assume the total number of movies on the global market is N and each of a group of K people has watched M movies, then the possibilities of the people's cultural preference of this group is $K \cdot N^M$. So the total possibilities of the global cultural preference diversity is :

$$D = \sum_{i=1}^{W} K_i \cdot N^{M_i}$$

where i enumerates the id of the group of people that watch the same number of movies, K represents the number of people within the group, and M denotes the number of movies that they watched within the group.

If the time-order becomes invariant for people's cultural preference, the total possibilities of the global cultural preference diversity becomes :

$$D = \sum_{i=1}^{W} K_i \cdot \frac{N^{M_i}}{N!}$$

Since the original formula of D with time-order-relevant formulation is now divided by N factorial, the diversity of the preference is greatly simplified. This makes it a lot easier for cultural product makers to capture people's future cultural preferences, and this partially explains why phenomenal cultural products such as Hollywood blockbuster could exist on global market.

What is more interesting is the conclusion drawn from the data-free Zeroshot recommender system algorithms, that is, **_human cultural consumption can be predicted fairly accurately individualistically without reference to past preferences_**. The conclusion can be safely derived from the observation that ZeroMat, DotMat, PoissonMat and PowerMat perform competitively with recommender systems with full historic data.

To better understand the conclusion, we consider, once again, the example of the phenomenal blockbusters. Since we discovered that human choice of cultural products is (mostly) history irrelevant, the highly personalized diversity of people's past preferences over cultural products is, therefore, not reflected on the global market. Therefore, the cultural product producers do not need to cater to customers' past preferences in details in order to produce a global blockbuster. The effect of history irrelevance of human culture choices has even a deeper impact on simplification of the diversity possibilities than time-order-invariance.

Another interesting phenomenon of the global movie market, is that repeating themes emerge once in a while. For example, superhero movies, war movies, etc. , they all have more or less something in common in respect of themes or story-lines, etc. However, the repeating elements of movies have not stopped the audience from entering the theatre. This phenomenon could also be explained by the history irrelevance property of human culture. Since the population is (mostly) memory-less, they would not grow tired of repeating elements even though they have experienced them many times before.

This interpretation may be generalized to a larger domain : During the human history, we've experienced through a series of cultural periods : The Renaissance, the Baroque, the Impressionism, etc. Each period can be categorized by common elements in fine arts, music, literature. The reason why repeating elements did not make people grow tired or sick for decades or even centuries, we believe, can also be explained (at least partially) by the history irrelevance property of human culture.

Another modern day example of history irrelevance property of human culture is popular music. There are many people who have listened to hundreds of hip-hop or rock music, without even feeling tired. Also, popular music with common elements are consistently used in TV series, and many people enjoy the music track. History irrelevance property of human culture might help explain this phenomenon.

**The Dark Side Story**

So what is the dark side story of the time-order-invariance and history-irrelevance properties of human culture ? Are we just, as in folklores, fish that bite the bait of psychological enticement with only a short memory of 7 seconds ? If the human culture is time-order-invariant and history-irrelevant, are we locked up in a static state where cultural elements today are just repeating themes that were already in existence thousands of years ago ?

To answer these questions, we first need to understand the underlying mechanism of Zeroshot learning algorithms. One common assumption of the Zeroshot learning algorithms including ZeroMat, DotMat, PoissonMat and PowerMat is the Pareto Law Distribution. All 4 algorithms use a special case of Pareto Law Distribution, namely Zipf Distribution, as its key property.

Researchers in 2021 published a paper pointing out that Zipf Distribution is a result of maximization of diversity and efficiency. Based on this conclusion, we infer that ***human culture is a quick optimization process for psychological indulgement.*** The Zipf phenomenon in cultural industry is exemplified in the fact that the user item rating values v.s the number of ratings follow Zipf distribution (or can be safely presumed to be so to generate competitive algorithmic results).

This optimization process completes its work pretty quickly. As we noticed on MovieLens and LDOS-CoMoDa datasets, or websites such as Douban, Goodreads and IMDB, the Zipf Distribution is in existence after only a short period of years (or even less) after the websites are created. However, we are cautious to claim this point, since human culture has evloved for thousands of years, and the websites are just loggers of human behavior. The optimization process might be completed a very long time ago in human history, and the website or open data are just evidence of the consequence.

***Because of the maximization of efficiency to acquire psychological indulgement, human culture has degenerated into a static lock state where history becomes (mostly) irrelevant.*** This is understanable, because if human culture choices are produced by human perceptions based on complicated computation and selection based on comparison between his past preferences and future choice candidate pool, it would be a lot time and energy consuming, and less efficient, than the history irrelevance process.

The dark side of the story is the Matthew effect of the global cultural market can jeopardize the overall healthy development of all mankind. With more and more global blockbusters dominating the major entertainment venues, human cultural choices will become more and more homogeneous. Most people will watch several movies in each year, and probably they'll be the same with other people. The lack of diversity will lead to concentration of technologies and financial resources, and it will kill creativity and emergence of new ideas or entertainment firms. The monopolistic tendency (or status quo) is theoretically provably bad for the economics and healthy development of the cultural industry.

**Cultural Policy**

To increase the selection pool of human culture choices and lift human culture from the static lock state, we recommend the governments implement the following cultural policies:

1. Establish investment funds for small entertainment companies
2. Provide high incentives for individual cultural product contributors
3. Compensate for less successful cultural products / firms
4. Promulgate laws to protect early-stage and small cultural producers
5. Encourage fair AI and fair technological practice
6. Raise awareness of the status quo of human culture

1. Establish investment funds for small entertainment companies

Investment funds is the powerhouse for internet and high-tech company's financial capabilities and needs. The rapid growth of internet business can be attributed to the full support of investment funds, who usually support startups financially without demanding paybacks. As we all know today, the most profitable cultural production firms are film and game companies. Due to the high cost of film and game making, it is rare that small business or individual contributors could make popular cultural products in this sector. Therefore, the need of financial support is necessary for the growth of minor players on the market.

The necessity of maintaining the presence of minor players and keeping the upward mobility possibilities for them guarantees the diversity of the global cultural market, and thus in some degree mitigates the history irrelevance effect and lift the human culture from the static lock state.

2. Provide high incentives for individual cultural product contributors

One important subsector of the cultural industry, especially for the intellectuals, is the book market. Book market is a perfectly competitive market without monopolistic structures on authorship. However, the cost of publishing books is high in many countries such as America and China. In most developed countries, the prices for books (particularly college textbooks) are also too high for normal readers.

The 2021 Chinese retail book market size is 98.68 billion RMB. This is a huge market. Although due to the COVID-19 pandemic, the market was still in recession. If the government could provide high incentives such as fee waivers or coupons to authors, the market could be boosted to grow even further. By introducing more authors with lower financial bars of entrance, we will be able to make the choice pool for human cultural consumption larger, and therefore in some degree mitigating the history irrelevance problem.

3. Compensate for less successful cultural products / firms

Countries like Japan, when taking off, usually provide compensation for embryonic industries. This is because such industries are perceived as future strategic industrial pillars. Government pay less emphasis on cultural industry than industries like 5G telecommunications and quantum computing. However, to salvage mankind from the memory-less and history-irrelevance state, government should pay equal attention to the cultural industry just like what they are doing to the strategic high-tech industries.

Special care should be paid to less successful cultural products / firms. One measure that global governance institutes have already been doing is the protection of intangible human heritage. Intangible human heritage is endangered human cultural relics that embody past human cultural preferences. It is a strong measure to rescue human cultural relics, and help mitigate the history irrelevance property of human culture, since intangible heritage itself is human history.

4. Promulgate laws to protect early-stage and small cultural producers

Intellectual property theft is a rampant problem in many developing nations, also it is a commonly encountered problem in developed nation as well. IP Laws and their effective execution serve as a protective moat around the castle of intangible assets of small cultural producers. They should be paid special care to encourage small cultural producers to introduce more diversity into the static history-irrelevant state.

Laws that force resource allocation biased towards small cultural producers should also be promulgated to boost quick growth of small firms. For example, land allocations or office allocations or loans should be allocated in favor of small cultural producers to promote fast introduction of diversified element to the static pool of cultural state.

Competition and privacy laws should also be enacted in a tolerating way that allow cultural producers leave their high posts in renowned companies to found startups so to increase the vibrancy in the market.

Another legal aspect of protective laws for small cultural producers is contract laws. Contract laws should make litigation easy and cheap for small cultural producers.

5. Encourage fair AI and fair technological practice

A modern day trend for cultural consumption is online purchase and view behavior. Chinese cultural consumers tend to use Douban as one of their major tools to find interesting book and movie reviews so they'll make purchases. IMDB is a western example. Websites as such usually have recommender systems so as to recommend interesting cultural products to customers.

One hot topic of recommender system is the fairness research. Fairness solves the popularity bias, selection bias, exposure bias, etc. With the awareness of the community and legal enforcement of governments, companies have started to make sure that less popular cultural items also receive enough exposure as measured by fairness metrics. The fair AI algorithms and other fair technologies should be promoted to counter the Matthew effect of the cultural consumption market.

6. Raise awareness of the status quo of human culture

In addition to the aforementioned 5 points, governments should also raise public awareness of the current situation of cultural industries. Public propaganda on the necessity of consistent monitoring of the health status of the cultural industry should be promulgated on a frequent basis. Besides, education on the urgency of the current problems related to cultural industry, especially the time-order invariance and history irrelevance properties.

**Social Impact**

Since the issues we discussed in previous sections in this paper exist not only in the cultural industry, but also other industries such as E-commerce, we extend our discussion to other social areas in this section.

The 6 algorithmic paradigms / approaches that we touched in the *Recommender Systems* section are also applicable in other websites such as Amazon and Alibaba. The time-order invariance and history irrelevance properties also exist for such e-commerce websites. This leads to the conclusion that human consumption behavior is also (mostly) history irrelevant, and this could explain the effect of phenomenal sales of a specific brand on the internet.

Policy suggestions for other social areas such as e-commerce websites are similar to our cultural policy suggestions, and the impact of such measures could never be overemphasized in the long term history of mankind.

**Conclusion and Future Work**

In this paper, we analyzed 6 algorithmic paradigms / approaches in the field of recommender systems, especially their performance and cultural implications on open datasets. We take 2 open datasets as our research benchmarks, namely MovieLens and LDOS-CoMoDa Datasets. The dataset contains human population samples of cultural ratings of the size tens of thousands or even millions. The algorithms that we analyze include classic algorithms and modern day new inventions.

The basic findings of our analysis of these algorithms' performance results on large-scale human cultural products are the time-order-invariance and history-irrelevance properties of the cultural products. In plain words, the consumers' future choice of cultural products is irrelevant of the order in which he consumed other products in the past, and also, we can predict the choice without using any historic data at all.

The 2 findings of us can help us explain some of the most interesting phenomena in human cultural history. For example, why is it possible to create global blockbusters ? Why repeating elements in fine arts and music do not make people feel tired. We gave several examples and interpretations based on our theory.

We also provide policy suggestions for regulation organizations and government agencies. We promote the idea that more diversity should be introduced in the current static locked-up human cultural industry. We gave up to 6 suggestion points that we hope could help mitigate the time-order-invariance and history-irrelevance problems.

We notice that our discussion can be extended to other social domain such as retailing online. The analysis is analogous to the cultural industry.

In future work, we'd like to quantify the history-irrelevance properties. In details, we'd like to explore the impact of the length of memory-span and how much it can effects the human cultural choice within the improvement space left over by algorithms such as ZeroMat and DotMat. We want to better understand exactly how much room has been left for us to improve the human cultural choice.

We would also like to explore the interpretability of Zipf Distribution's efficacy in modeling the human cultural consumption process. We want to create a quantifiable connection between the degree of psychological indulgement and Zipf Distribution. Possibly physiological sensors will be built to monitor people's cultural enjoyment and measurements on correlation between the experience and consumers' past consumption records will be studied.

We believe our work in this paper is one of the pioneering research studies in a field what we'd like to call *Computational Cultural Studies*. This area should explore large-scale datasets acquired by modern day websites or other venues. Data mining algorithms and artificial intelligence approaches serve as major research tools in this area. We hope our research has pointed out the right direction for future work in the area.